\documentstyle[preprint,aps,prl,psfig]{revtex}
\tighten
\draft
\begin{document}

\preprint{\today}
\title{Aging in a Structural Glass\footnote{Talk given at STATPHYS
20, Paris, July 20-24 1998}}
\author{Walter Kob}
\address{Institut f\"ur Physik, Johannes Gutenberg-Universit\"at,
Staudinger Weg 7, D-55099 Mainz, Germany}
\author{Jean-Louis Barrat}
\address{ D\'epartement de Physique des Mat\'eriaux \\ Universit\'e
Claude Bernard and CNRS, 69622 Villeurbanne Cedex, France}

\maketitle

\begin{abstract}
We discuss the relaxation dynamics of a simple structural glass which
has been quenched below its glass transition temperature. We
demonstrate that time correlation functions show strong aging effects
and investigate in what way the fluctuation dissipation theorem is
violated.
\end{abstract}


\noindent

\section{Introduction}
Whenever a system whose relaxation time is large is driven out of
equilibrium, it can be expected that its dynamics shows aging effects.
This means that observables that in equilibrium are constant
become time dependent and time correlation function that in
equilibrium depend only on time {\it differences} will now depend on {\it two}
times. Typical examples for such situations are ferromagnetic
coarsening or the relaxation dynamics of spin and structural
glasses~\cite{aging_theo}. The investigation of such aging phenomena is by
no means a new subject~\cite{struik78_mckenna89}, but due to new
theoretical approaches~\cite{aging_theo}, which have led to a variety of
predictions that call for being tested, this field has recently become a
very active area of research.

For the case of structural glasses not much is known about the aging
dynamics {\it on a microscopic level}, since the experiments needed to
address these questions are unfortunately quite difficult. This is in
contrast to computer simulations, since these easily allow to study the
system on a microscopic level and thus give access to all observables
of interest. The price for this advantage is that only relatively small
time scales and systems can be studied, but it has turned out that
these disadvantages are not too serious. In such simulations one usually
mimics the experimental setup in that the system is prepared in an
equilibrium state and at time zero driven out of equilibrium, e.g. by
decreasing the temperature or by applying an external field.
Subsequently the system is allowed to relax for a certain waiting time
$t_w$ and then one starts to measure its properties, such as the
density, the magnetization or a time correlation function. This
approach is also the one that we will use in the present work in an
attempt of gain a better understanding of the dynamics of structural
glasses at low temperatures. 

\section{Model and Details of the Simulation}

For the investigation of aging effects it is useful to be able to
change the waiting time over as many decades as possible and, for a
given waiting time, to study the subsequent relaxation dynamics over a
long time.  Therefore it is advisable to study aging phenomena for
models that are simple enough to be simulated over a large
time window and are still reasonably realistic to catch the essential
features of structural glasses. One such model is a binary
Lennard-Jones mixture whose dynamical properties in its strongly
supercooled state have been investigated in great
detail~\cite{kob_lj}.  In these studies it has been shown that the
dynamics of this system can be described well by means of mode-coupling
theory~\cite{mct}, with a critical temperature $T_c$ around 0.435 (in
reduced energy units).

The particles in this 80:20 mixture interact via a Lennard-Jones
potential of the form $V_{\alpha\beta}(r)=4\epsilon_{\alpha\beta}
[(\sigma_{\alpha\beta}/r)^{12}-(\sigma_{\alpha\beta} /r)^{6}]$ where
$\alpha$ and $\beta$ denote the type of particles (which we call ``$A$''
and ``$B$''). In the following we will use $\sigma_{AA}$ and
$\epsilon_{AA}$ as the unit of length and energy, and
$(m\sigma_{AA}^2/48\epsilon_{AA})^{1/2}$ as the unit of time, where $m$
is the mass of the particles, which is independent of the species.  The
parameters of the potential are $\epsilon_{AA}=1.0$, $\sigma_{AA}=1.0$,
$\epsilon_{AB}=1.5$, $\sigma_{AB}=0.8$, $\epsilon_{BB}=0.5$, and
$\sigma_{BB}=0.88$. The total number of particles was 1000 and in order
to minimize finite size effects we used a cubic box of size 9.4 and
periodic boundary conditions. The equations of motions have been
integrated with the velocity form of the Verlet algorithm with a step
size of 0.02. Most of the runs were 5 million steps long.

In order to have aging phenomena a non-equilibrium situation has to
be generated, which was done as follows. Starting from an equilibrium
configuration at a high temperature ($T_i=5.0$) we quenched the system
at time $t=0$ to a final temperature $T_f<T_c$. This was done by coupling
the system periodically (every 50 time steps) to a heat bath. The
system was allowed to evolve at the temperature $T_f$ for the waiting
time $t_w$ and subsequently we started the measurement of the time
correlation functions. In order to improve the statistics of the result
this procedure was repeated 6-10 times for different initial
conditions.

\section{Results}

\begin{figure}[f]
\psfig{file=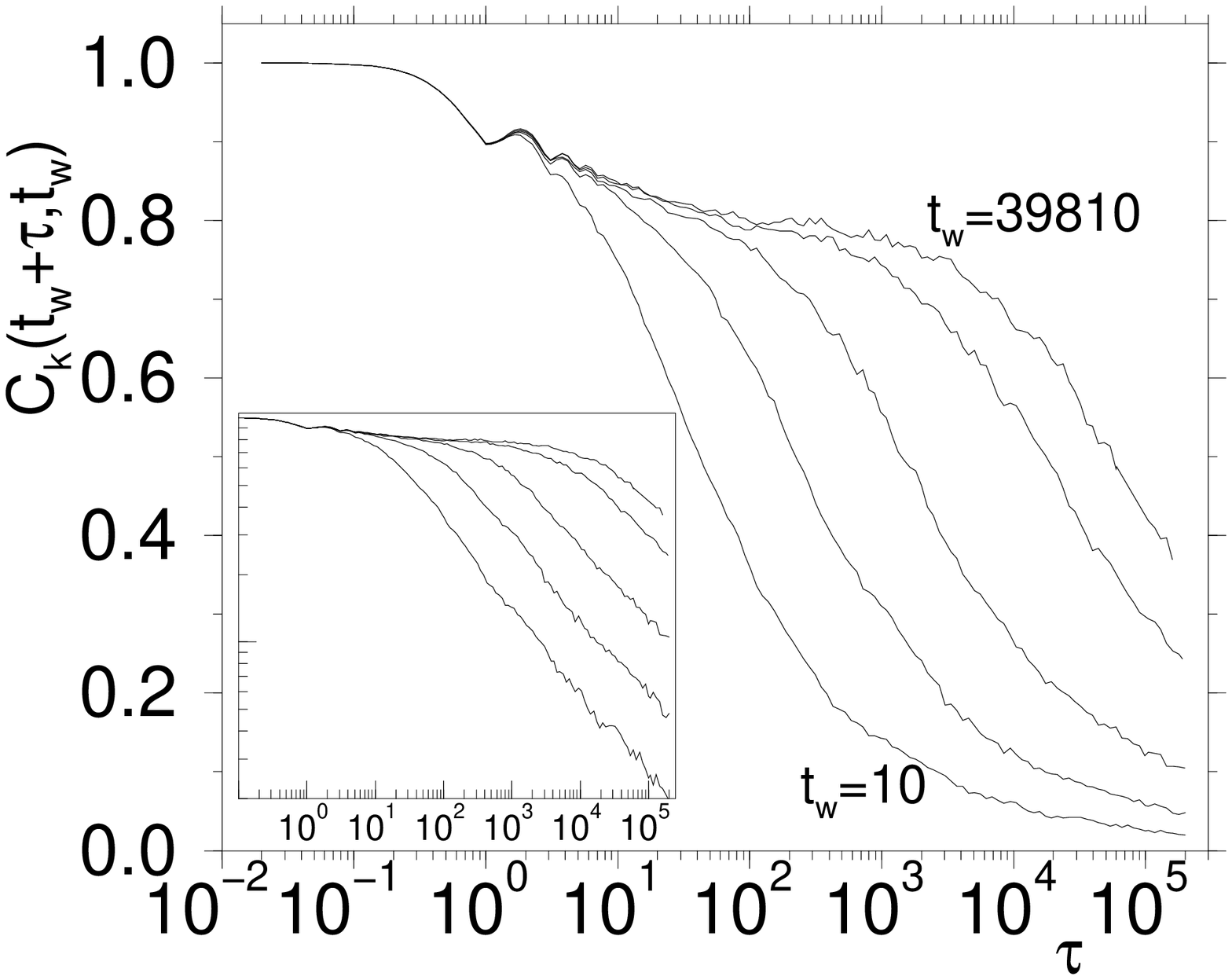,width=13cm,height=9.5cm}
\caption{Time dependence of $C_k(t_w+\tau,t_w)$ for different waiting
times ($t_w=10$, 40, 1000, 10000, 39810). $T_f=0.4$. Inset: The same 
correlation functions in a double logarithmic representation.}
\label{fig1}
\end{figure}

As shown previously in Ref.~\cite{kob97} quantities that, {\it in
equilibrium}, do not depend on time, such as the total energy of the
system, are not very sensitive to the aging process. Much more
pronounced non-equilibrium effects are observed for time dependent
quantities, such as the intermediate scattering
function~\cite{kob97,barrat98} or the mean squared
displacement~\cite{parisi97}.  In the following we will therefore study
the time and $t_w$ dependence of $C_k(t_w+\tau,t_w)$, the
generalization of the self intermediate scattering function to
non-equilibrium situations. This observable is defined by
\begin{equation}
C_k(t_w+\tau,t_w)= {1\over N} \sum_j \langle \exp\left[i{\bf k}\cdot
\left({\bf r}_j(t_w+\tau)-{\bf r}_j(t_w)\right)\right] \rangle,
\label{eq1}
\end{equation}
where ${\bf k}$ is the wave-vector and ${\bf r}_j(t)$ is the position
of particle $j$ at time $t$. In Fig.~\ref{fig1} we show the time
dependence of $C_k(t_w+\tau,t_w)$ for different waiting times (see
figure caption). The value of $k=|{\bf k}|$ is 7.2, the location of the
maximum in the structure factor, and $T_f=0.4$, i.e. a temperature which
is only 10\% below $T_c$.  The main figure shows $C_k$ in a log-lin
representation. We see that at short times the curves do not depend on
$t_w$, i.e.  no aging effects are observed~\cite{footnote}.  For longer
times we find, however, very pronounced aging effects in that a curve
with a finite waiting time $t_w$ starts to leave the common curve
observed at short times and decays towards zero. In Ref.~\cite{kob97}
it was shown that the time at which this pealing off from the envelope
curve occurs is on the order of $t_w$.

In the inset we show the same data in a log-log plot. From this figure
it becomes evident that at long times the relaxation of $C_k$ is
described well by a power-law with an exponent that is independent of
$t_w$ and which is around 0.4. Qualitatively similar results are found for
other values of $k$. For short times we have found that the approach to
the plateau is described well by a power-law, $C_k(t_w+\tau,t_w)
\propto \tau^{-a}$, with an exponent around 0.45, a time dependence
that is compatible with the prediction of mean-field theories of
aging~\cite{aging_theo}.

\begin{figure}[h]
\psfig{file=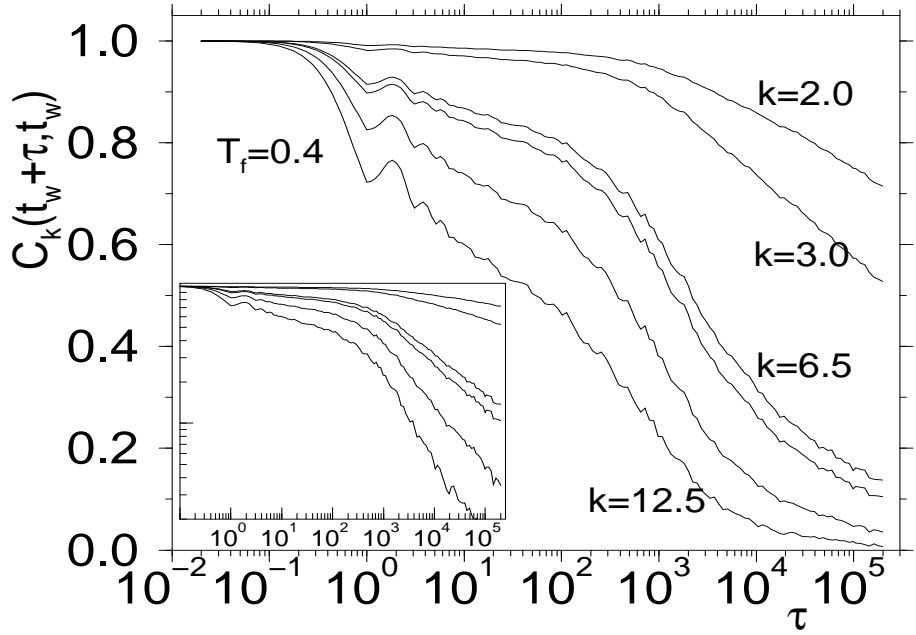,width=13cm,height=9.5cm}
\caption{Time dependence of $C_k(t_w+\tau,t_w)$ for different
wave-vectors $k$ and $t_w=1000$. $T_f=0.4$. From top to bottom: 
$k=2.0$, 3.0, 6.5, 7.2, 9.6, 12.5. Inset: The same
correlation functions in a double logarithmic representation.}
\label{fig2}
\end{figure}

In Fig.~\ref{fig2} we show $C_k(t_w+\tau,t_w)$ for $t_w=1000$ for
different values of $k$. We see that the relaxation of the curves slows
down dramatically when $k$ is decreased. For example, the curve for
$k=3.0$ takes about 100 times longer to decay to 0.5 than the curve for
$k=6.5$. This factor has to be compared with the one expected for a
diffusive process which is $(6.5/3.0)^2 \approx 4.7$, i.e. much
smaller. The inset shows the same correlators in a log-log
representation. From it we recognize that the time dependence of the
curves is compatible with a power-law and that the exponent decreases
significantly with decreasing wave-vector. (We note that for small
values of $k$ the time dependence is also compatible with a logarithmic
decay, a law that is proposed from domain growth
models~\cite{fisher88}.)

\begin{figure}[h]
\psfig{file=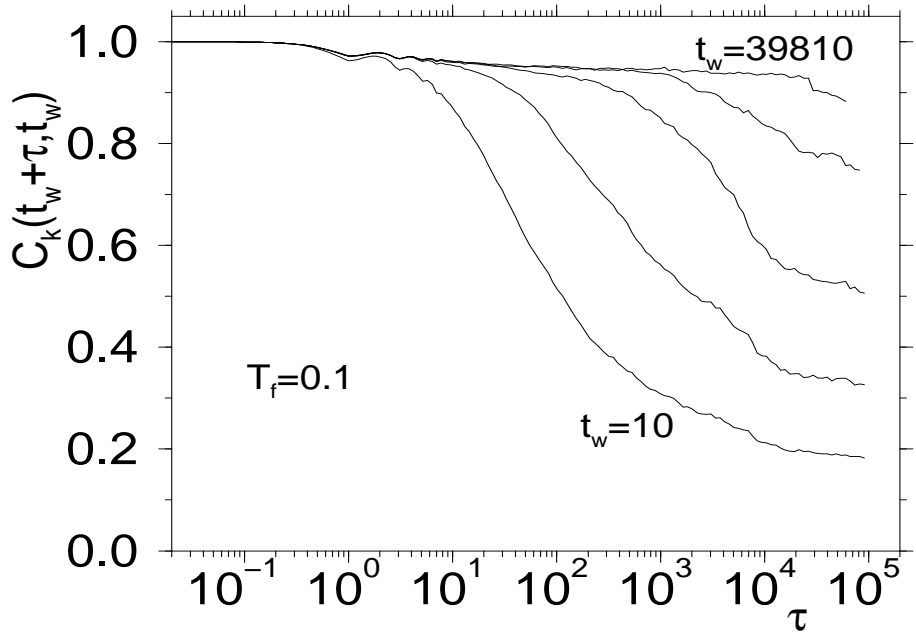,width=13cm,height=9.5cm}
\caption{Time dependence of $C_k(t_w+\tau,t_w)$ for different waiting
times ($t_w=10$, 40, 1000, 10000, 39810). $T_f=0.1$.}
\label{fig3}
\end{figure}

The results presented so far have been for the final temperature
$T_f=0.4$, i.e. a temperature that is only about 10\% below the
critical mode-coupling temperature of the system
($T_c=0.435$)~\cite{kob_lj}. If the final temperature is significantly
lower than $T_c$ the relaxation behavior is qualitatively different. In
Fig.~\ref{fig3} we show $C_k(t_w+\tau,t_w)$ for the same waiting times
and the same value of $k$ as in Fig.~\ref{fig1}, but now for $T_f=0.1$.
A comparison of the two figures shows that the short time dynamics is
qualitatively similar, except that $q_{EA}$, the height of the plateau
at intermediate times which is called nonergodicity- or
Edwards-Andersen parameter, has increased with decreasing $T_f$. Such a
$T_f$ dependence can easily be understood by recalling that at low
temperatures this height is related to the amplitude of the vibrations
of the particles in their cages and that within the harmonic
approximation this amplitude is expected to be proportional to the
temperature. In fact, a closer inspection of the figures shows that
$1-q_{EA}$ is indeed proportional to $T_f$.

The main difference between the relaxation behavior for $T_f=0.4$ and
the one for $T_f=0.1$ occurs for long waiting times in the time
regime in which the correlation functions decay below the plateau. The
early stage of this decay for $T_f=0.1$ is qualitatively similar to the
one for $T_f=0.4$. However, for very long times, $\tau > 10^4$, the
correlators for $T_f=0.1$ seem to show an additional plateau, a feature
which is not present in the correlators for $T_f=0.4$. A closer
inspection of the curves for the {\it individual} samples (for
$T_f=0.1$ we have 9 different samples) revealed that the reason for
this second plateau is given by a quite dramatic (0.1-0.2) and fast
decay of the correlation function shortly before the plateau. The time
at which this decay occurs depends on the sample but is usually on the
order of $10^3-10^4$ time units. An analysis of the motion of the particles
in the time range at which this sudden drop occurs shows that the
decay is related to a very collective movement in which on the order of
10\% of the particles move by about 0.1-0.5 units of length in one
direction. This observation can be rationalized as follows. After the
quench the configuration of the particles is very unfavorable and thus
the system relaxes very quickly. If the system is given a bit more
time, i.e.  for larger waiting times, it has enough time to relax to a
state which is no longer that unfavorable (for the given $T_f$) and
hence does not relax that quickly. For intermediate and large $t_w$
it will hence explore {\it for short times $\tau$} only than part of the
configuration space which corresponds to the motion of the particles
within their cages. However, the system will locally still have quite
large stress fields and, given enough time, will yield to these
stresses and hence show a rupture like motion which is the reason for
the fast drop in $C_k$. Since this type of motion is so abrupt it is
unlikely that a mean-field like theory will be able to give a correct
description of it, except perhaps in a phenomenological way.  (We note
that this situation is reminiscent to the one of the mode-coupling
theory of supercooled liquids, since also in that case the so-called
``hopping processes'' strongly affect at low temperatures the very
continuous, flow-like motion of the particles~\cite{mct}.)

\begin{figure}[h]
\psfig{file=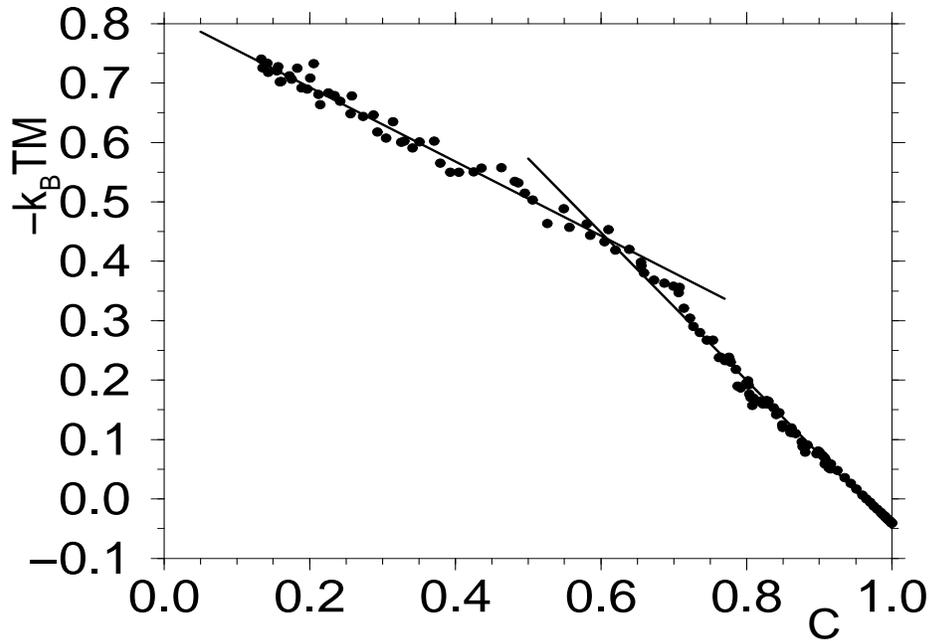,width=13cm,height=9.5cm}
\caption{Parametric plot of $-k_B T M$ versus $C$, where $M$ is the
integrated response. $T_f=0.4$, $t_w=1000$. The two straight lines have
slopes around $-0.62$ and $-1.0$}
\label{fig4}
\end{figure}

A very interesting result of the theories of aging is related to the
violation of the fluctuation dissipation theorem (FDT). In
equilibrium the autocorrelation function $C_A(t)$ 
of an observable $A$ is related to the response $R_A(t)$ of $A$ to its
conjugate field by the FDT, i.e. $R_A(t)=-(1/k_BT) \partial
C(t)/\partial t$. For the non-equilibrium situation this relation is
no longer valid but it can be generalized to
\begin{equation}
R_A(t',t)=\frac{1}{k_BT} X_A(t',t) \frac{\partial C_A(t',t)}{\partial t},
\label{eq2}
\end{equation}
where $t'\geq t$ and $X_A(t',t)\leq 1 $ is defined by this equation.
Hence $T/X_A(t',t)$ can be considered as the temperature for which the
usual connection between the time correlation function and the response
holds~\cite{cugliandolo97}. The concept of such a temperature has been
used in the glass literature for a long time, in the form of the
so-called
``fictive temperature'', but has remained so far a ill defined
quantity. In contrast to this the definition given by Eq.~(\ref{eq2}) is
from a theoretical {\it and practical} point of view much clearer and
useful and hence more appealing. Instead of calculating the response
$R_A(t',t)$ directly, where now $A$ is the one-particle density
distribution, we proceeded (basically) as follows~\cite{footnote2}.
After having quenched the system at time $t=0$ we let it relax for a
time $t_w$.  At time $t_w$ we applied a sinusoidal field with
wave-vector $k$ and amplitude $V_0=0.3$ which coupled to the density and
calculated the expectation value of the density
distribution\cite{footnote3}.  Therefore we obtain the integrated
response $M(t_w+\tau,t_w)$
\begin{equation}
V_0 M(t_w+\tau,t_w)  =  V_0 \int_{t_w}^{t_w+\tau} R(t_w+\tau,t) dt
\quad .
\label{eq3}
\end{equation}
It has been argued that for $t_w$ and $\tau$ large
$X_k(t_w+\tau,t_w)$ becomes a function of $C_k$ only, i.e. 
$X_k(t_w+\tau,t_w)=x(C(t_w+\tau,t_w))$, where $x$ is a function of one
variable~\cite{aging_theo}. Using this and Eq.~(\ref{eq3}) we obtain
\begin{equation}
M(C)=\frac{1}{k_BT}\int_C^1 x(c) dc,
\label{eq4}
\end{equation}
where we used the fact that $C_k(t_w+\tau,t_w)=1$ for $\tau=0$. This
result suggests that a parametric plot of $k_BT M$ versus $C$ is a
useful way to look at the data and in Fig.~\ref{fig4} we show such a
plot.  For large values of $C$, which corresponds to short times, we
see that $M(C)$ is essentially a straight line with slope close to
$-1.0$. This means that $x(C)$ is close to $-1$, i.e.  that the FDT
holds. With decreasing $C$, corresponding to increasing time $\tau$,
the curve is compatible with a straight line with slope $-m>-1$.
Therefore we find that in that region $-x(C)=m<1$ and hence that the
FDT is violated.  Note that a linear dependence of $M$ on $C$ in the
non-FDT region has also been found of ``$p$-spin''
models~\cite{aging_theo} and thus give support to the
hypotheses~\cite{kirkpatrick89_parisi97b} that structural glasses are
in the same universality class as such models.  Finally we mention that
the $T_f$ dependence of the slope $m$ is essentially linear. In
particular we find for $T_f=0.1$ $m=0.1$ and for $T_f=0.3$ and
$T_f=0.4$ $m=0.45$ and $m=0.62$, respectively. Assuming a linear
dependence on $T_f$ we thus expect for $T_f=T_c$ a value around 0.7,
i.e. significantly smaller than 1.0, as might be expected from
mean-filed theory.

\section{Summary}
We have presented some results of a large scale computer simulation in
which the non-equilibrium relaxation dynamics of a simple structural 
glass was investigated by quenching the system below its glass
transition temperature. We find that time correlation functions, such
as the generalization of the self intermediate scattering function,
show a very strong dependence on the waiting time and thus are useful
observables to study the aging properties of the system. We have also
calculated the response function in order to investigate the
violation of the FDT and have found that for short times FDT holds
whereas for long time the theorem is violated. The temperature
dependence of the FDT-violation factor $X$ is similar to the one
found for certain spin glass model, thus giving evidence that at low
temperatures the phase space structure of such systems is similar to
the one of structural glasses. Finally we mention that many of the
aging phenomena discussed here have already been observed in computer
simulations and experiments on spin glasses~\cite{aging_theo} thus
giving further evidence for this point of view.

Acknowledgements: We thank L. Cugliandolo, J. Kurchan, A. Latz, and G.
Parisi for many useful discussions. Part of this work was supported by
the Pole Scientifique de Mod\'elisation Num\'erique at ENS-Lyon and 
the Deutsche Forschungsgemeinschaft through SFB 262.

\clearpage
\newpage

\begin{references}
\bibitem{footnote}
Note that the small amplitude oscillations in $C_k$ at short times 
stem from the periodic coupling of the system to the heat bath.
\bibitem{footnote2}
More details on the calculation of the response can be found in
Ref.~\protect\cite{barrat98}.
\bibitem{footnote3}
We checked for the linearity of the response by varying $V_0$.
\end{references}
\end{document}